\documentclass[aps,prb,showkeys,preprint,showpacs,12pt]{revtex4-1}
\usepackage{amssymb}
\usepackage{hyperref}
\usepackage{graphics,graphicx}
\usepackage{dcolumn}
\usepackage{amsbsy}
\usepackage{amsmath}
\usepackage{epsfig}
\usepackage{color}
\usepackage{textcomp}

\begin{document}
 \title{Enhanced magnetic anisotropy of Nickel nanosheet prepared in Na-4 mica}
\author{Sreemanta Mitra$^{1,2}$}
\email[]{sreemanta85@gmail.com}
\author{Amrita Mandal$^{1,2}$}
\author{Anindya Datta $^{3}$}
\author{Sourish Banerjee$^{2}$}
\author{Dipankar Chakravorty$^{1,\dag}$}
\email[]{mlsdc@iacs.res.in}
\affiliation{
$^{1}$
 MLS Prof.of Physics' Unit,Indian Association for the Cultivation of Science, Kolkata-700032, India.\\ }
\affiliation{
$^{2}$
Department of Physics, University of Calcutta, Kolkata-700009, India.\\}

\affiliation{
$^{3}$
University School of Basic and Applied Science (USBAS),Guru Govind Singh Indraprastha University,New Delhi, India\\}

\begin{abstract}
 Nanosheets of nickel with thickness equal to 0.6 nm have been grown within the interlayer spaces of Na-4 mica.
The sheets are made up of percolative clusters of nanodisks. Magnetization characteristics indicate a superparamagnetic
behavior with a blocking temperature of 428 K.The magnetic anisotropy constant as extracted from the coercivity data has been found to be higher than that of bulk nickel
by two orders of magnitude. This is ascribed to a large aspect ratio of the nickel nanophase. The Bloch exponent is also found to be considerably different from that of
bulk nickel because of a size effect. The Bloch Equation is still found to be valid for the two dimensional structures.
\end{abstract}
\maketitle

\section{Introduction}
The magnetic properties of two dimensional nanostructures with large aspect ratio have
 received considerable attention in recent years\cite{epjb1}. Two-dimensional systems display magnetic properties 
distinct from those of bulk materials, and it remains a challenge to understand their behavior \cite{hornbook}. 
The magnetic properties of nanoparticles depend strongly on the particle size and shape and the results 
are interesting not only from the point of view of basic physics but also of technological applications
such as ferrofluids, magnetic refrigaration etc.      \cite{jms1,pss1,jms2,poppjmmm,rajjmmm}. 
Though the synthesis of two dimensional transition metal having fcc structure has been rarely explored, 
two dimensional single crystalline nickel has been grown recently by a solution phase method \cite{lengnt,lengjpcc}. 
These nanosheets are triangular and hexagonal in shape having a thickness of 6 nm and edge length of 15.4 nm. 
\par
We had previously synthesized two dimensional silver \cite{bosejncs},manganese \cite{mitrajpcc}
and compounds like $BaTiO_{3}$ \cite{danjmc}
 and GaN \cite{santanujpd,santanuapl} 
by a template assisted method. In order to synthesize them, sodium fluorophlogopite, commonly known as 
Na-4 mica, 
was used as the template. Na-4 mica is a layered material.
The interlayer spaces have been exploited to grow two dimensional nickel. The thickness of such regions was 
estimated from the interplanar spacings of the concerned planes as calculated from the
X-ray diffraction pattern of the material \cite{kodamajmc1}. 
From the latter the basal plane (001) spacing
of the hydrated Na-4 mica was found to be 1.21nm. Subtracting the interplanar spacing of (002) planes viz; 0.61nm \cite{kodamajmc1}
from the above we obtained 0.6nm, which we take as a measure of the interlayer thickness for hydrated Na-4 mica.
These spaces have unusually large number of four cations per unit cell, loosely bound
with the upper and lower block oxygen ions via weak electrostatic interactions.  
The cations can be exchanged with suitable ions. After proper treatment the formation of desired materials
within the nanodimensional space can be achieved. Due to this small thickness, much higher aspect ratio 
can be induced in the system than that reported till date. We had recently used Na-4 mica to grow nickel nanosheets. The study
was restricted to delineating the magnetodielectric effect in the system \cite{selfepl} which was found to be substantial.
In this present investigation we have explored the magnetic anisotropy behaviour of two dimensional metallic nickel
of thickness 0.6 nm synthesized within the nanochannels of Na-4 mica.  
\section{Experiment}
  \subsection{Synthesis}
      Synthesis of Na-4 mica has been described elsewhere \cite{mitrajpcc,danjmc,pkmprb}. The Na-4 mica powder was kept inside
 a saturated solution of nickel nitrate (as obtained from E.Merck (India) Ltd.) in water for two and a half months,
 at 333K in an air oven, for the ion exchange reaction, $2Na^{+}\Leftrightarrow Ni^{2+}$ to occur. The pH of the solution was kept at 7.0. 
The concentration of $Ni^{2+}$ ions in the solution was $4.2 (10^{21} /cc)$. The mixture was stirred at regular interval. 
Alongside the ion exchange process, the nanospaces were filled with $Ni^{2+}$ cations and $(NO_{3})^{-}$ anions. A longer duration
 and slightly higher temperature than that required for ion-exchange process [8] were used because we wanted to ensure
 the introduction of $Ni^{2+}$ and $(NO_{3})^{-}$ ions as well within the nanospace. The powder was then dried and washed 
thoroughly with deionized water several times, so that no $Ni(NO_{3})_{2}$ molecule remained on the surface of the powdered sample.
 Finally, the sample was subjected to a reduction treatment. The ion exchanged powder, in an alumina boat, was brought to 1273 K
 in nitrogen atmosphere inside a tubular furnace and then it was subjected to Hydrogen $(H_{2})$ gas flow for 1 h.
 Hydrogen reduced the $Ni^{2+}$  ions in the nanospaces to form metallic nickel. The sample was brought back to room temperature
 by furnace cooling in nitrogen atmosphere. The temperature and time of reduction had to be determined by a trial and error method.
 The conversion to metallic nickel was possible only under this optimum condition.
  \subsection{Characterization}
 The phases present in the composites synthesized were identified by X-ray diffraction studies on them. A Bruker D8 XRD SWAX diffractometer
was used with Cu $K_{\alpha}$ radiation. The diffraction angle was varied from $10^{o}$ to $80^{o}$. For microstructural studies
an etching technique was used for separating the nickel nanosheets from Na-4 mica - the etchant being a 40{\%} HF aqueous solution. 
The nickel phase was retrieved from the solution by centrifuging the latter in a SORVALL RC 90 ultracentrifuge at 
40,000 rpm for 30 minutes. After washing the sample in water several times for the removal of HF it was dispersed in acetone(AR Grade, E-Merck, India).
A JEOL 2010 transmission electron microscope operated at 200 kV was used to study the microstrure of the sample.
\\
Magnetic properties were investigated by a Superconducting Quantum Interference Device (SQUID) magnetometer (Quantum Design MPMS XL)
in the temperature range 2 to 300 K with an applied field up to 30 kOe.
\section{Results and Discussion}
    \subsection{Structure Analysis}
        
   The X-ray diffractogram obtained from Na-4 mica powder synthesized in this work is shown in figure 1. The d-spacings calculated from 
 the diffraction angles pertaining to different peaks match well with those reported in the
 literature \cite{parkcm,leejmc}. Figure 2(a) shows the X-ray diffraction pattern of the composite comprising Na-4 mica and nickel.
 The characteristic diffraction peaks for both the phases with their corresponding lattice planes are marked on the figure. 
 It may be noted that the XRD reflection (002) is strongly widened. This is explained as arising due to the partial pillaring of the
 mica structure during the growth of nickel phase at high temperature. Figures 2(b),2(c),2(d) show  amplified views of the diffraction peaks
 corresponding to (111), (200), (220) planes of nickel phase respectively.
 The data in the figure were fitted by a Gaussian function as indicated by the solid line, which was then employed to 
  calculate the Full Width Half Maxima (FWHM) values. 
 The latter are summarized in table 1. It is evident that the values are different from each other showing thereby the anisotropic
 features of the grown nickel phase. 

\begin{figure}
\includegraphics[width=8.5cm]{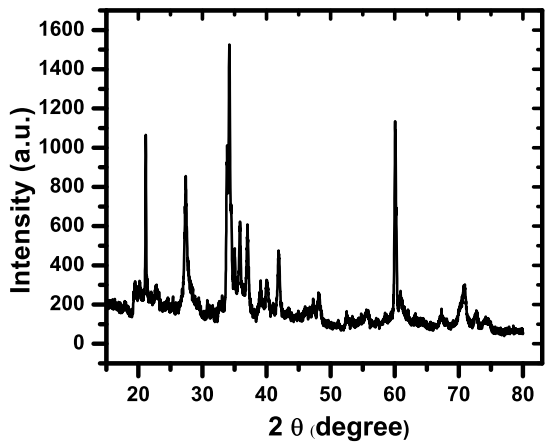}
\caption{%
  XRD pattern of synthesized Na-4 mica.}
\label{fig:1}
\end{figure}
\begin{figure}
\includegraphics*[width=8.5cm]{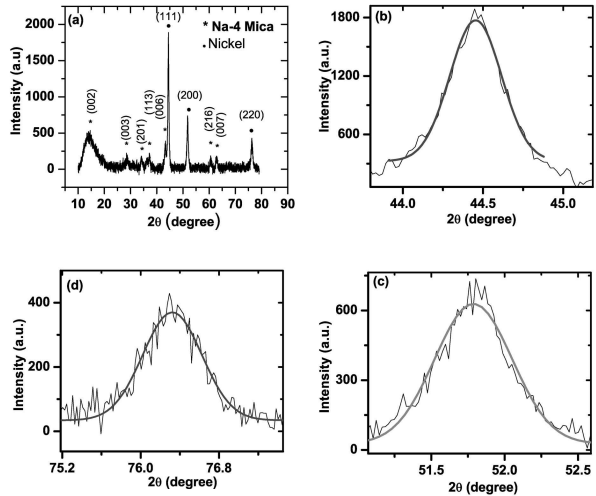}
\caption{%
        (a) XRD pattern of Nickel-Na-4 mica composite.(b),(c),(d) show the amplified views of 
        diffraction peaks corresponding to (111), (200), (220), of nickel respectively.
        The solid line being the fitting with a Gaussian function.}
\label{fig:2}
\end{figure}
\begin{table}
\centering
  \caption{FWHM for the Nickel phase.}
  \begin{tabular}{cc}
    \hline
   Plane & FWHM (in degree) \\
    \hline
    (111) & 0.38 \\
    (200) & 0.62 \\
    (220) & 0.71 \\
    \hline
  \end{tabular}
\end{table}
We have also tried to delineate the thickness of the nickel nanosheet by dispersing the extracted nickel
 entities (see section 2.2) on a freshly cleaved mica (SPI Supplies, West Chester, PA, USA) 
 surface mounted on measuring platform of an Atomic Force Microscope (AFM) (Vecco Model,CP- II). 
The changes in AFM stylus position as it is drawn over the samples whose images are given in 3(a) . 
  The green line in the z-color scale is the base line, denoting the substrate (mica) surface from which the heights were measured.
 The big, bright particles are unreacted flakes of Na-4 mica.
It is to be noted that the steps in the line profiles arose due to the low resolution of the imaging, and definitely not real. 
 The thickness of a single sheet of nickel was found  to be 0.6 nm. However, we have also obtained areas of thickness 1.2 nm and 1.8 nm, 
because due to this method of sample preparation one may obtain a situation
 where few nickel nanosheets juxtaposed on top of each other. In table 2 we have given a quantitative analysis of the occurence of the various thicknesses as obtained 
from different line profile plots of figure 3(a). Two such line plots have been shown in the figure 3(b).The line profiles are done on two lines (viz; line1 and line2).
 The statistical observations have been summarized in figure 4(b). Figure 4(a) shows the 3-dimesional view of figure 3(a).   
Due to low magnification of the AFM image it has not been possible to determine the nanodisk shape and size. However, from the line profiles by considering 
the heights of 0.6 nm we measured the base width thereof and could conclude that the nanodisk diameters varied from 10 nm to 30 nm. To substantiate this 
 transmission electron micrograph of nickel nanosheets grown within Na-4 mica has been shown in figure 5(a). It can be seen that the
 nickel nanosheets have been formed by the percolation of nickel nanodisks of diameter varying from 8 to 26 nm. Figure 5(b) gives the high resolution
 transmission electron micrograph of one of the nanosheets. The interplanar spacing obtained from the image is 0.20 nm, which is in agreement with the
 $d_{hkl}$ value for the (111) plane in nickel. Since nickel has an fcc crystal structure the formation of (111) plane is thermodynamically favorable, and
 this has been verified by the high resolution lattice image.  The presence of nickel phase is further confirmed by figure 4(c), which shows the selected area
 electron diffraction pattern.  The $d_{hkl}$ values are calculated from figure 5(c) and summarized in table 3. These are compared with 
Joint Committee on Powder Diffraction Standards (JCPDS) data (file no.04-0850) for nickel and Na-4 mica\cite{parkcm} respectively.
\begin{figure}
\includegraphics*[width=15cm]{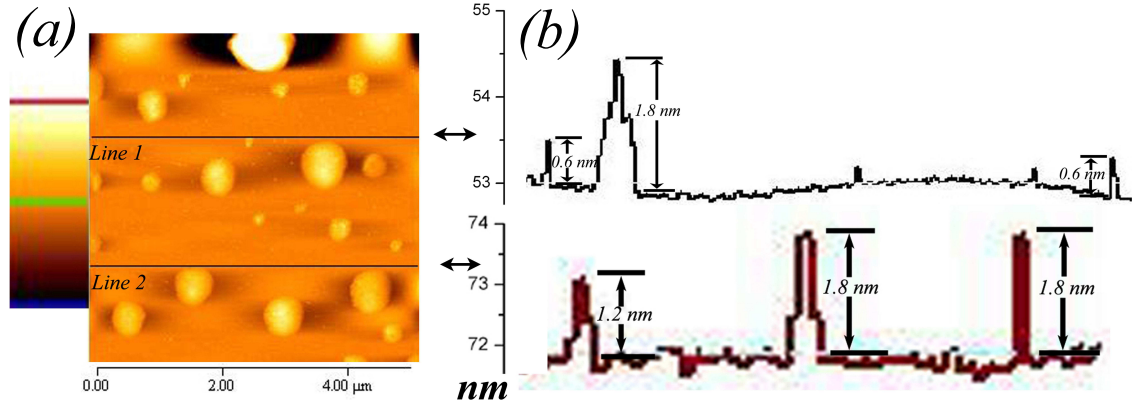}
\caption{%
  (a) AFM image of a portion of the disperson on a freshly cleaved Mica surface. The green line in the z-color scale denotes the base line from which the height profile 
was measured. The black color represents the depth below the base signifying broken portion of mica substrate.
  (b) The typical height profiles corresponding to the lines in figure 3(a). }
\label{fig:3}
\end{figure}
\begin{figure}
\includegraphics*[width=15cm]{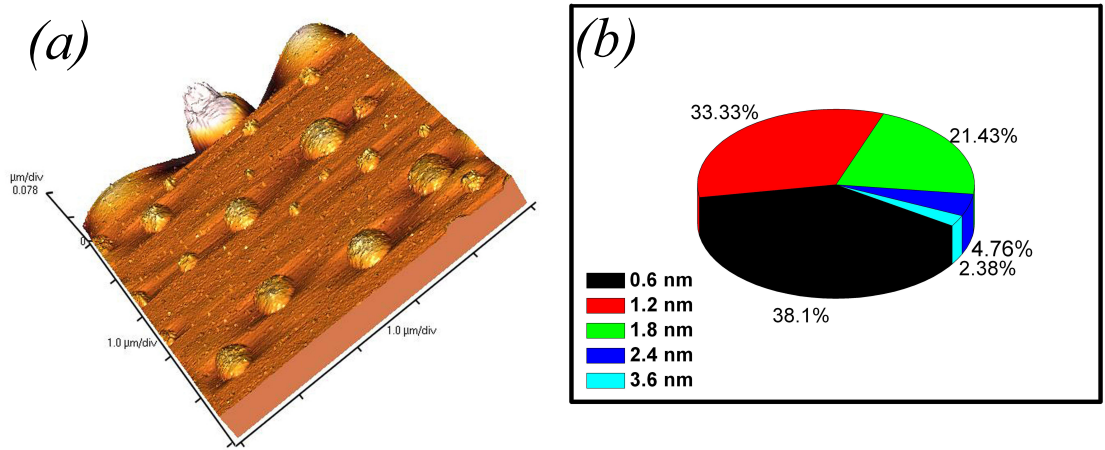}
\caption{%
  (a)The 3 dimensional view of the figure 3(a).
  (b)The percentages of obtaining thickness of intergral multiples of 0.6 nm (shown in a Pie plot) as obtained from fig. 3(a).}
\label{fig:4}
\end{figure}
\begin{table}
\centering  
\caption{Number of observations for a particular thickness as obtained from figure 3(a). }
   
  \begin{tabular}{cc}
    \hline
   Thickness measured & No. of times observed\\ (nm) & \\
    \hline
    0.6 & 16 \\
    1.2 & 14 \\
    1.8 & 09 \\
    2.4 & 02 \\
    3.6 & 01 \\
   \hline
    
  \end{tabular}
  \label{table2}
\end{table}

\begin{figure}
\includegraphics*[width=8.5cm]{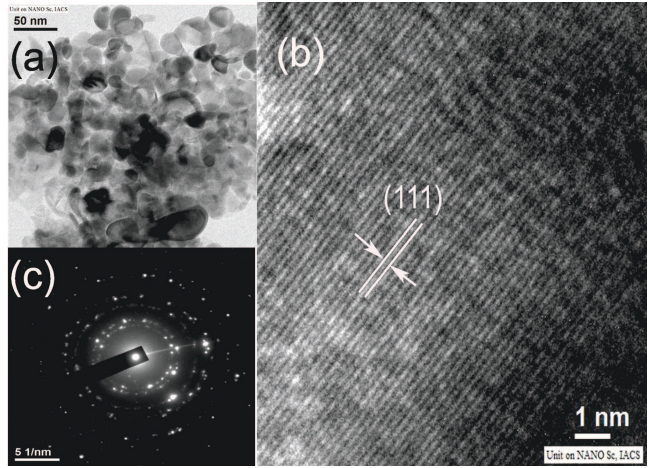}
\caption{%
     (a) Random assembly of nickel nanosheets. 
     (b) High resolution lattice image from a nickel nanosheet.
     (c) Selected area electron diffraction pattern of (a)}
\label{fig:5}
\end{figure}
\begin{table}
\centering
  \caption{Comparison of interplannar spacings $d_{hkl}$ obtained 
           from electron diffraction pattern of nanocomposite
            with standard JCPDS data.}

  \begin{tabular}{ccc}
    \hline
   Observed & Na-4 mica & Nickel\\ (nm) & (nm) & (nm)\\
    \hline
    0.25  & 0.244    &  $-$ \\
    0.20  & $-$      & 0.203 \\
    0.15  & 0.16     & $-$ \\
    0.12  & $-$      & 0.124 \\
    0.10  & $-$      & 0.101  \\
    0.09  & $-$      & 0.088 \\
    \hline
$-$ indicates that there exist
\\
 no planes with these interplanar spacings
\\
 in the phases concerned.

  \end{tabular}

  \label{table3}
\end{table}
\subsection{Magnetic Properties}
    The variation of magnetization as a function of temperature under
 both field cooled (FC) and zero field cooled (ZFC) conditions and measured at magnetic fields 50 Oe 
 and 500 Oe  are shown in figures 6(a) and 6(b) respectively.
The magnetization values were normalized in terms of the composite mass of the nanocomposite system. The diamagnetic contribution \cite{mitrajpcc}
from the Na-4 mica powder was duly subtracted from the measured values.
 The figures in the inset of figure 6(a) and 6(b) show those data in the higher temperature side.
 From this, it is evident that the ZFC and FC data do not merge with each other at 300 K,indicating that the spins are still blocked even at room temperature.
  The data indicate superparamagnetic behaviour above a certain temperature.
 Since at both the applied fields the FC curve deviates from the ZFC curve at the highest temperature measured (300 K),
 the superparamagnetic blocking temperature of the nanosheets is above 300 K. This high blocking
 temperature is a signature of their
 highly anisotropic nature. This is borne out by magnetization-magnetic field hysteresis curve
 shown in figure 7. 
In the temperature range of measurement of the M-H isotherm we observe a hysteretic feature which signifies that the material is ferromagnetic at the room temperature.
 It may be noted here that due to lack of data at temperatures above 300 K it
 was not possible to determine the exact value of blocking temperature $T_{B}$ from FC ZFC curves. 
       We have measured the coercivity at different temperatures, from the M-H isotherms, and figure 8 represents the
 variation of coercivity as a function of temperature. The experimental data were fitted to the relation \cite{childressapl}
\begin{equation}
 H_{C}(T)=H_{c}(0)[1-(\frac{T}{T_{B}})^{1/2}]
\end{equation}
where, $T_{B}$ is the blocking temperature, and $H_{C}(0)$ is the coercivity at 0 K. From the  least square fitting of the
 experimental data the extracted value of the blocking
temeperature was found to be 428 K. It may be noted that eq.1 pertains to spherical particles. 
We have fitted our results to this equation and the calculated value of \textquoteleft $\kappa$\textquoteright (see below) refers to 
the effective anisotropy constant for an equivalent set of spherical particles having the same volume as those of the nanodiscs in the present system.
The magnetic anisotropy constant \textquoteleft$ \kappa$\textquoteright  has been calculated using equation \cite{cullitybook}
\begin{equation}
 \kappa=25\frac{k_{B}T_{B}}{V}
\end{equation}
\begin{figure}
\includegraphics*[width=8.5cm]{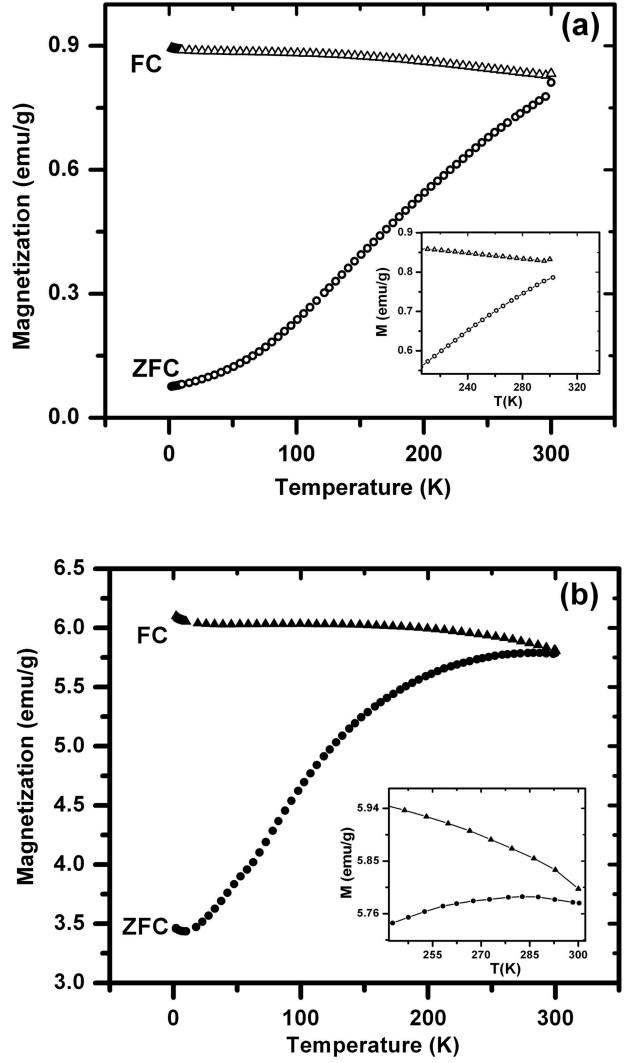}
\caption{%
  Variation of magnetization with temperature under 
  both field cooled (FC) and zero field cooled (ZFC) 
  conditions measured at (a) 50 Oe and (b) 500 Oe.
  The insets show the magnified views near maximum temperature of measurement.
  The open circle is the ZFC and open 
  triangle is the FC data (50 Oe).The closed circle
  and triangle are the ZFC and FC datas for the 500 Oe respectively.}
\label{fig:6}
\end{figure}

\begin{figure}
\includegraphics*[width=8.5cm]{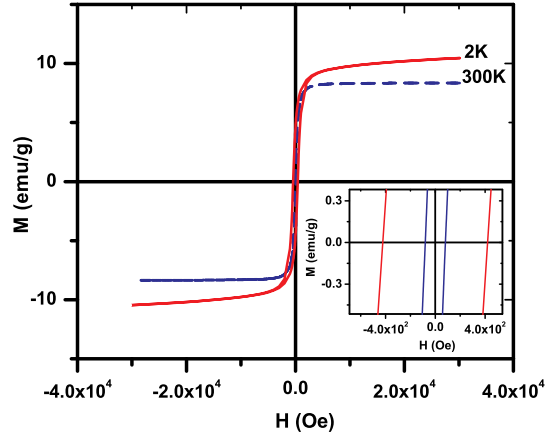}
\caption{%
  Magnetization-magnetic field isotherm curves measured at 2 K and 300 K. 
The inset shows the expanded view at the lower magnetization showing the presence of finite coercivity.}
\label{fig:7}
\end{figure}

\begin{figure}
\includegraphics*[width=8.5cm]{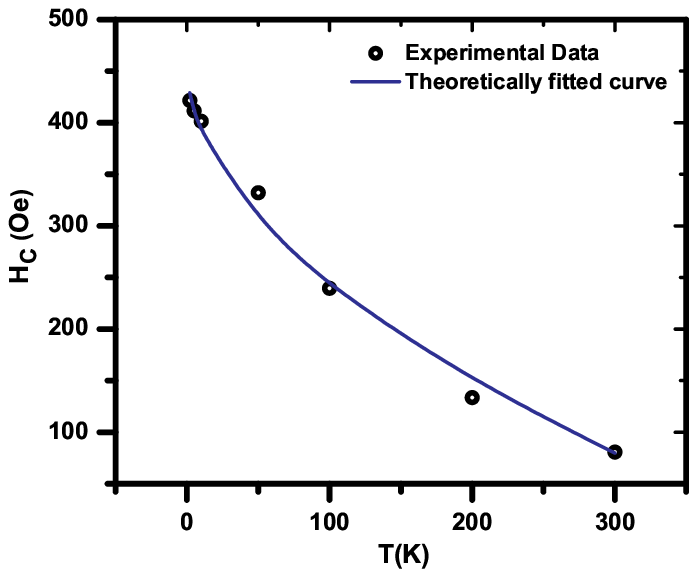}
\caption{%
 Variation of coercivity with temperature and its theoretically fiited curve.}
\label{fig:8}
\end{figure}
where $k_{B}$ is Boltzman constant $T_{B}$ is blocking temperature extracted from $H_{C}-T$ curve, and V is the volume 
of a single particle as measured from TEM image. The anisotropy constant was calculated to be $7.77\times10^{6} erg/cm^{3}$, 
which is about two orders of magnitude higher than that of bulk nickel, viz; $5.0\times10^{4} erg/cm^{3}$ \cite{bonderjmmm}. 
This high magnetic anisotropy constant can be attributed to the highly anisotropic shape of the nickel nanosheets.
As a matter of fact the room temperaure coercivity is also two orders of magnitude higher than that of bulk nickel, viz;0.7 Oe \cite{chemmat1}.
It should be pointed out here that the magnetic measurements were carried out on a powdered sample with the nanosheets aligned 
at random with respect to the magnetic field direction. Hence, the magnetic anisotropy as extracted above can not be assigned to 
any specific direction but can be described as arising due to an integrated effect of all directions of the sample system.
In case of spheres, nanorods, and nanowires the measured coercivities at 5 K are 200 Oe, 630 Oe, and 730 Oe respectively \cite{jmmm310}.The 
value of 415 Oe obtained in our system seems consistent with these because our samples are two dimensional in nanoscale.
In figure 9 the variation of saturation magnetization as a function of temperature has been shown.
For achieving saturation magnetization from the M-H isotherms, we have plotted [M-(1/H)] and $M_{s}$ was obtained by the
extrapolation of the data to $\frac{1}{H}  \rightarrow 0$.    
It is observed that the magnetization increases as the temperature is lowered due to a decrease in 
thermal energy\cite{wumrb} . It is to be noted, that the saturation magnetization obtained in our studies is around 12 emu/g.
 This is much lower than that observed in pure fcc nickel at 300 K \cite{hwangjmr}. This difference is caused by the fact that our
 measurements were carried out on the composite powder in which Ni nanosheets formed a  fraction of the material.
 Na-4 mica occupied the major volume of the sample. As a result the true magnetization can be obtained by dividing the
 value by a fraction \textquoteleft f \textquoteright delineating the value of the nickel nanosheets in our composite system.  
We have fitted the magnetic data with Bloch Equation \cite{kodamajmmm,zhangprb}
\begin{equation}
 M_{S}(T)=M_{s}(0)[1-BT^{b}]
\end{equation}
\begin{figure}
\includegraphics*[width=8.5cm]{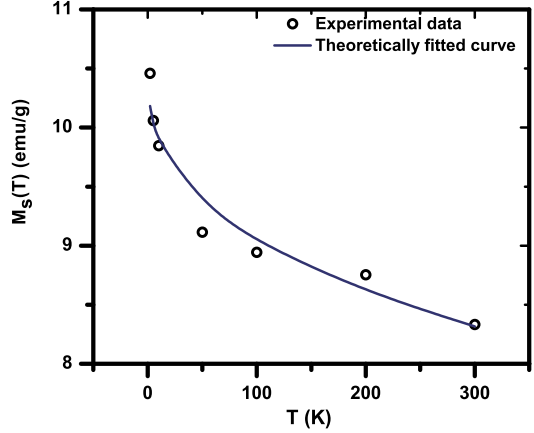}
\caption{%
 Variation of saturation Magnetization with temperature with the theoretically fiited curve.}
\label{fig:9}
\end{figure}
The values of B and b were found to be $0.031 K^{-1}$ and 0.34 respectively.
 It can be noted that the value of Bloch exponent\textquoteleft b\textquoteright is smaller than 1.5. 
 Such a discrepancy has been adduced to an interfacial or a size reduction effect \cite{zhangprb}.
The value of \textquoteleft b \textquoteright was reported to be lower than 1.5 as the particle size was reduced \cite{zhangprb}. Our data are consistent
with this finding reported earlier. However, the comparison has to be done carefully because our system is two dimensional. As a result,
the nature of variation in the value of the Bloch coefficient is opposite to that reported in case of spherical iron nanoparticles.
 In the present system, both the aspects become operative. There is a large interface
 between the nickel sheets and the Na-4 mica blocks. 
 Also the sizes of nickel nanodisks have values in the range 8 to 26 nm,with a thickness of 0.6 nm and as a
 result of reduction in size, the Bloch exponent shows a value smaller than 1.5 .  However, Bloch equation is still valid for the two dimensional systems.
\section{Conclusion}
 Nanosheets of nickel with thickness equal to 0.6 nm have been grown within the nanosized 
interlayer spaces of Na-4 mica. The sheets are made up of percolative clusters of nanodisks. 
Magnetization characteristics indicate a superparamagnetic behavior with a blocking temperature of 428K. 
The magnetic anisotropy as extracted from the coercivity data,is found to be higher than that of bulk nickel
by two orders of magnitude.
 This is due to a large aspect ratio of the nickel nanophase. The Bloch exponent
 also deviates considerably from that of bulk nickel because of a size effect. 
The enhancement of coercivity in these nanofilms as compared to that in bulk nickel will make the material useful
as a memory device.
\section*{Acknowledgement}
 The work was supported by  Department of Science and Technology,Govt.of India, New Delhi
 under an Indo-Australian Project on Nanocomposites. 
Sreemanta Mitra and Amrita Mandal thank University Grants Commission for Senior Research Fellowships. D.Chakravorty
 thanks Indian National Science Academy for awarding an Honorary Scientist's position.

%
\end{document}